# Scattering optical elements: stand-alone optical elements exploiting multiple light scattering

**Jongchan Park[1], Joong-Yeon Cho[2], Chunghyun Park[1,3], KyeoReh Lee[1], Heon Lee[2*], Yong-Hoon Cho[1,3*], YongKeun Park[1*]**

[1]Department of Physics, Korea Advanced Institute of Science and Technology, Daejeon 305-701, Republic of Korea.

[2]Department of Materials Science and Engineering, Korea University, Anam-Dong 501, Sungbuk-gu, Seoul 136-713, Republic of Korea

[3]KI for the NanoCentury, Korea Advanced Institute of Science and Technology, Daejeon 305-701, Republic of Korea

*Correspondence:
Prof. YongKeun Park,
Department of Physics, Korea Advanced Institute of Science and Technology, Daejeon 305-701, Republic of Korea. Tel: (82) 42-350-2514, Email: yk.park@kaist.ac.kr

Prof. Yong-Hoon Cho
Department of Physics, Korea Advanced Institute of Science and Technology, Daejeon 305-701, Republic of Korea. Tel: (82) 42-350-2549, Email: yhc@kaist.ac.kr

Prof. Heon Lee
Department of Materials Science and Engineering, Korea University, Anam-Dong 501, Sungbuk-gu, Seoul 136-713, Republic of Korea, Tel: (82) 2-3290-3284, Email: heonlee@korea.ac.kr

**Optical design and fabrication techniques are crucial for making optical elements. From conventional lenses to diffractive optical elements[1,2], and to recent metasurfaces[3,4,5], various types of optical elements have been proposed to manipulate light where optical materials are fabricated into desired structures. Here, we propose a scattering optical element (SOE) which exploits multiple light scattering and wavefront shaping. Instead of fabricating optical materials, the SOE consists of a disordered medium and a photopolymer-based wavefront recorder, with shapes the wavefront of impinging light on demand. With the proposed stand-alone SOEs, we experimentally demonstrate control of various properties of light, including intensity, polarisation, spectral frequency and near field. Due to the tremendous freedom brought about by disordered media[6], the proposed approach will provide unexplored routes to manipulate arbitrary optical fields in stand-alone optical elements.**

Classical refractive optical elements such as lenses use the laws of refraction to manipulate light (Fig. 1a). Light is considered as a ray, and its propagation is mainly governed by the shapes of optical materials. With classical optical elements, it is challenging to generate complex optical fields. Alternatively, a diffractive optical element (DOE) exploits the wave nature of light (Fig. 1b)[2]. With microscopic structures fabricated on a transparent substrate, the DOE is a wavefront transformer that generates phase delays for desired optical functions[1,2]. However, the usage of DOEs is still limited to specific applications because they require complicated optical design and microfabrication techniques.

Recently, various types of metasurfaces have been developed, and these have significantly extended the applicability of optical elements[3,4,5]. Subwavelength structures in the metasurfaces allow full optical manipulation in the classical limit and demonstrate unprecedented optical functions. However, difficulties in optical design and fabrication become more significant, which together with economic aspects limits the practical applications of the metasurfaces.

Here, we present a method to generate optical elements exploiting multiple light scattering. This technique, referred to as scattering optical element (SOE), serves as a general optical wavefront transformer, which does not require complicated optical design or fabrication processes.

The SOE consists of a disordered turbid layer and a holographic photopolymer film (Fig. 1c). When an arbitrary beam is impinged upon, randomly distributed speckle patterns are formed after passing turbid media due to multiple light scattering. However, when an optimal wavefront is found and impinged upon, a desired optical field can be obtained after the turbid media[6]. Recently, this concept of coherent control of light propagation in complex media has been exploited in various wavefront shaping techniques using a spatial light modulator (SLM)[6,7].

The principle of SOE is to find, record, and reconstruct an optimal wavefront to a holographic film in order to make a stand-alone optical element exploiting multiple scattering. Importantly, the SOE utilises multiple light scattering in turbid media which can be exploited to access optical parameters such polarisation[8], spectrum[9], spatiotemporal control[10], plasmonics[11], optical nonlinearity[12], optical near fields[13] and negative refractive index[14]. It can be understood that versatile, stand-alone optical elements can be made by shaping the optical wavefront instead of fabricating the optical materials.

The optimal incident wavefront can be found by several existing methods[15]: iterative optimisation methods using an SLM[7,16], or measuring the transmission matrix of the turbid media[17,18,19], or it can also be determined by time-reversal approaches using optical phase conjugation[20,21,22]. Once the optimal wavefront is found, it can be recorded on a holographic film in order to be used as a stand-alone optical element. To record and reconstruct an optimal wavefront for generating a desired optical field through a turbid layer, a holographic photopolymer film (Bayfol®HX, Covestro, Germany) was used. A volume phase grating that generates a desired optical field is formed on the photopolymer film[23]. Once the volume phase grating is formed, the photopolymer generates a permanent and stable wavefront, which makes both the photopolymer film and the scattering layer into a stand-alone optical element.

The SOE is experimentally realised in a three-step procedure: wavefront optimisation, recording and reconstruction (Fig. 2). During the wavefront optimisation step, the wavefront of the incident light is optimised by using an SLM based on an iterative algorithm used in Ref.[7] (Supplementary section B). In the recording process, the beam optimised by the SLM, which serves as a signal beam, and a reference beam simultaneously illuminate the photopolymer to form a volume hologram. Finally, the desired optical field is reconstructed and impinged upon the turbid medium by illuminating the photopolymer film with the reference beam. Once the desired wavefront is recorded on the photopolymer, the SLM is no more required, and the SOE element can be used independently. The use of the photopolymer film substitutes the role of a bulky SLM to constitute the SOE.

To validate the proposed idea, we experimentally generated an optical focus using the SOE and quantitatively analyzed the operating performance. Before the wavefront optimisation, a random speckle pattern is formed behind the SOE (Fig. 3a). After the optimisation using $N = 212$ optical modes, a sharp focus is generated (Fig. 3b). The focus size was approximately 1.5 μm (FWHM), which was theoretically determined by the effective numerical aperture (NA) that the SOE supports and the NA of an imaging system. A peak-to-background intensity ratio of the focus was measured as $117 \pm 8$, which is comparable to the theoretical limit[7], $\pi(N - 1)/4 + 1 \cong 165$. With the use of a larger number of controllable modes $N$, the contrast can be enhanced, but at the expense of longer optimisation time.

In practice, photopolymer films undergo volume shrinkage upon polymerisation during the recording process, which may result in the imperfect reconstruction of the wavefront. We observed that this occurs in the SOE;

however, it does not significantly degrade the performance quality of the SOE. The focus was clearly reconstructed behind the scattering layer using the photopolymer, and a resultant peak-to-background ratio of the focus was measured as 88 ± 9 (Fig. 3c).

In the SOE, the precise alignment of the optimal wavefront to the corresponding turbid layer is essential for generating a desired optical field. To investigate the stability of the SOE, we measured the intensity of the generated focus while mechanically translating the photopolymer film in a lateral direction (Fig. 3d). As expected, the focus intensity decreases as the mismatch of the incident wavefront increases. Although the performance of the wavefront reconstruction using the SOE is highly sensitive to the alignment of the photopolymer film and the turbid layer, once the components are fixed to constitute an optical element, they remain highly stable for a long time. Figure 3e presents the peak-to-background ratio of the reconstructed focus as a function of time. The correlation value of the measured intensity patterns between the initial time and after $10^5$ s remains as 0.98 (Supplementary Fig. S3).

To further demonstrate the capability of the SOE, images of a micrometer-sized letter 'K' were generated. To find the optimal wavefront corresponding to the letter, a transmission matrix of the turbid layer was measured using a wavefront shaping method[19] (Supplementary section B). The letter 'K' is clearly shown when 2000 optical modes are used (Fig. 4a). When a fewer number of optical modes were employed, the optimisation process time can be reduced inversely proportional to the number of used optical modes, but it will also degrade the quality of reconstructed images because of decreased peak-to-background ratio.

The complex forms of an optical field carrying high-spatial-frequency components can be constructed using the SOE regardless of the number of optical modes in the incident wavefront. However, the use of lower optical modes will result in undesirable background speckle noise. To quantitatively analyze the aforementioned effect, the ratio of total intensity at the points of interest to the average intensity of the background speckle is measured as a function of the number of optical modes used for wavefront optimisation (Fig. 4b).

To further demonstrate the applicability of the SOE, we demonstrate the control of diverse optical parameters in generating optical foci, including polarisation states, wavelengths and near-field components. Due to the high degree of freedom in multiple light scattering, desired optical information can be readily accessed by shaping the wavefront of an incident light to a complex medium (Figs. 5a-c). Using the SOE, two foci were generated which have different polarization states (Fig. 5d) or spectral (Fig. 5e) components, respectively. In addition, the use of random nanoparticles as a turbid medium allows access to near-field information of scattered light[13, 24, 25]. To demonstrate the near-field manipulation using the SOE, scanning optical microscopy is used to detect optical near-field intensity at the proximity of the turbid medium (Supplementary section E). Figure 5h clearly shows the formation of subwavelength focus using the SOE. The FWHM of the reconstructed focus was approximately 130 nm, corresponding to λ/4.5. This result indicates the direct control of optical near fields using the SOE. This result clearly demonstrates the control of various optical parameters - which have been previously shown using an SLM[8, 9, 13] - can be realised in an independent optical element.

In summary, we present a method to make versatile stand-alone optical elements exploiting multiple light scattering. In contrast to existing technology, the present method does not need complicated optical design and fabrication processes, nor the use of a bulky and expensive SLM. Instead, the optimal wavefront is found, permanently recorded onto the holographic photopolymer, and then reconstructed on demand as a stand-alone optical element. Importantly, the use of multiple light scattering enables versatile optical field control ranging from focusing to the control of various optical parameters, as demonstrated in wavelength, polarisation and near field. In addition, multiplexed holographic recording and reading may also enable the various controls of optical fields on demand. Furthermore, the present method can be readily scalable to larger optical elements.

## Methods

**Experimental setup.** A coherent He-Ne laser ($\lambda$ = 633 nm, 21 mW, HNL210L, Thorlabs, USA) and a DPSS laser ($\lambda$ = 532 nm, 100 mW, Shanghai Dream Lasers Technology, China) were used for illumination. The beam was divided into a sample and a reference arm using a beam splitter. The sample beam was expanded by a 4-$f$ telescopic system and directed to a phase-only SLM (LCOS-SLM, X10468, Hamamatsu, Japan). The diffracted beam from the SLM was projected onto a turbid medium, which consisted of two layers of holographic diffusers (∅1″ Circle Pattern Diffuser, ED1-C50, Thorlabs, USA). The transmitted light was collected by an objective lens (×60, NA = 0.7, LUCPLFL, Olympus, USA) and imaged through a 4-$f$ telescopic imaging system, with a lateral magnification of ×25. The image was recorded by a CCD camera (LT365R, Lumenera, USA).

During the optimisation process, a blazed grating phase pattern was displayed on the SLM simultaneously to separate spatially the modulated beam. An iris was placed in the Fourier plane of the SLM in order to filter out only the spatially modulated beam. The other reference beam was coupled into a single mode optical fiber. During the optimisation process, the reference beam was blocked by using a shutter. The power of the beam at the surface of the photopolymer was set to 35 $\mu W/cm^2$ by using a neutral density filter in order to avoid photoreactions of the photopolymer film. The exposure of the camera was set to 200 ms because of a low-intensity level during the optimisation. It takes 1.6 s for finding each optimal phase value for an optical mode. After the optimisation, the total power of the signal and the reference beams are set to 350 $\mu W/cm^2$. After an exposure of 20 s, the photopolymer is left dark for 1 min and then cured by illuminating it with a UV light ($\lambda$ = 365 nm, 120 m, WM365L2-C1, Thorlabs, USA) for 15 min.

To demonstrate optical near-field control, a modified near-field scanning optical microscopy (NSOM) (Alpha SNOM, WITec, Germany) was used. The detailed experimental setup and procedure can be found in Ref.[13]. The beam diffracted from the SLM was demagnified by ×1/600 and imaged onto the surface of a turbid medium which consisted of $ZrO_2$ nanoparticles. Near-field components of a transmitted beam were collected by using an NSOM probe and detected by a photomultiplier tube. The photopolymer film and the surface of the turbid medium were placed on the conjugated image planes of the SLM. Thus, when optimal wavefront is reconstructed using the photopolymer film, the image of the wavefront is directly projected onto the surface of the turbid medium.

## ACKNOWLEDGMENTS
This work was supported by KAIST, and the National Research Foundation of Korea (2015R1A3A2066550, 2014K1A3A1A09063027, 2013M3C1A3063046, 2014M3C1A3052537, 2012-M3C1A1-048860) and Innopolis Foundation (A2015DD126). The authors acknowledge the support by Bayer MaterialScience AG for providing the photopolymer film

## Author Contributions
J. P. performed the experiments and analysed the data. J.-Y. C, C. P., K. L., H. L. and Y.-H. P. contributed experimental and analytic tools. Y. P. conceived and supervised the project. J. P. and Y. P wrote the manuscript.

**Author Information** Supplementary information is available in the online version of the paper. Reprints and permissions information are available online at www.nature.com/reprints. Correspondence and requests for materials should be addressed to Y.P. (yk.park@kaist.ac.kr).

## Competing financial interests
The authors declare no competing financial interests.

**Figures with captions**

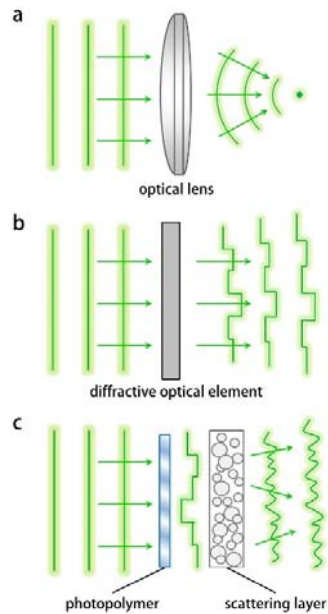

**Figure 1 | Principle of scattering optical element. a**, Classical (geometric) optical element uses laws of refraction and reflection to manipulate light. **b**, Diffractive optical element utilises the wave nature of light. It modifies the spatial phase profile of the light into a desired form. **c**, Scattering optical element (SOE) uses a turbid layer which scrambles optical information as a part of the optical element. With the aid of the high degree of freedom inherent in the turbid layer, the SOE can address the optical information such as polarisation states, spectral frequency, high-spatial frequency and near-field components in the manner of wavefront shaping.

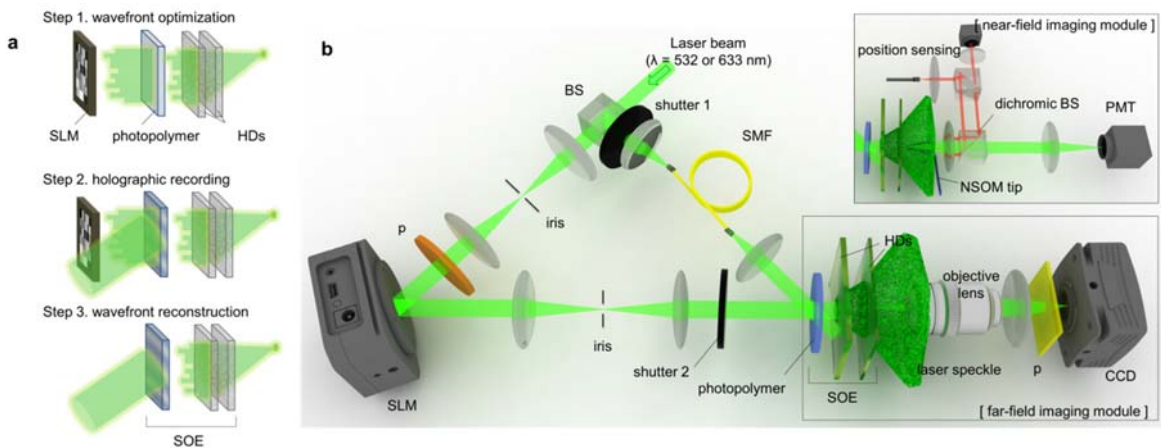

**Figure 2 | Experimental scheme. a,** Experimental procedure. (1) Optimisation process. The spatial phase profile of the beam incident on a scattering medium is modulated by the SLM to generate the desired optical field behind the turbid layer. The intensity of the beam is set to below the photo-reaction exposure of the photopolymer film during the optimisation process by using a neutral density filter. (2) Recording process. Both the optimised signal beam and the reference beam illuminate the photopolymer film to form volume phase grating on it. After

construction of the volume grating, the photopolymer is exposed to ultraviolet light for curing. (3) Reconstruction process. The desired optical field is generated by illuminating the photopolymer using the reference beam. **b,** Experimental setup. A coherent laser beam is divided by a beam splitter to serves as a signal and a reference beam. The reference beam is coupled to a single mode optical fiber. A wavefront of the signal beam is optimised by the SLM and is passed through a photopolymer film and a turbid layer. The intensity signal is measured by a charge-coupled device (CCD). HD: holographic diffuser, BS: beam splitter, P: polariser, SLM: spatial light modulator, SMF: single mode optical fiber, PMT: photomultiplier tube.

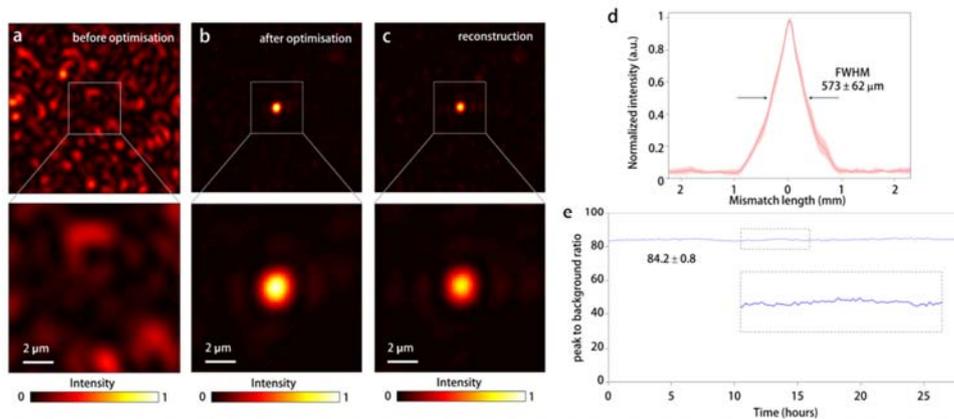

**Figure 3 | Optical focusing using the scattering optical element**. Intensity of the optical field behind a scattering medium before (**a**) and after (**b**) optimisation. **c**, Optimised field is reconstructed using scattering optical element. FWHM of the focus was measured as ~ 1.5 μm, which was decided by the effective NA of the SOE. **d**, Intensity of the focus was measured as a function of a radial mismatch length of the wavefront. When the mismatch length is the same size as the segment (which is the case when the correlation between the optimised wavefront and the actually impinging wavefront on the turbid medium becomes zero) the focus totally disappears. **e**, Time-persistence of the focus. SOE is highly stable over a long time. The measured correlation value of the intensity pattern at $t = 0$ s and $t = 10^5$ s was 0.98 (see Supplementary Fig. S3).

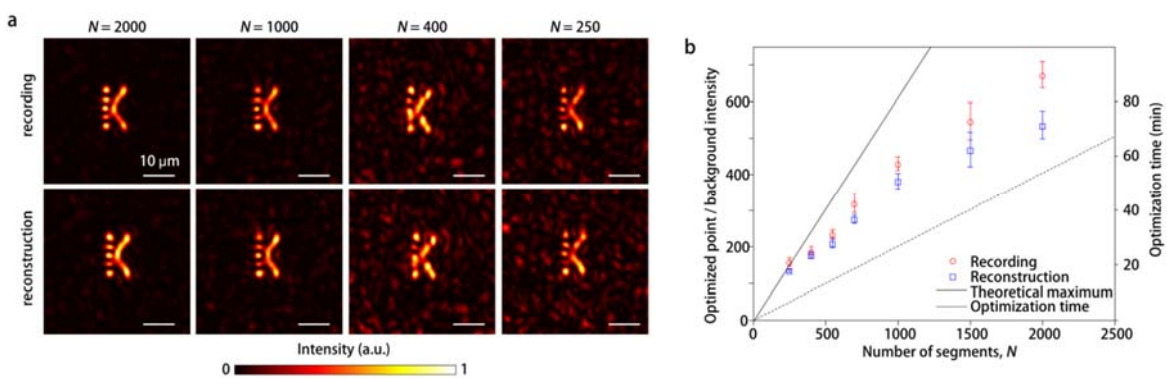

**Figure 4 | Micrometer-sized two-dimensional image formation using the scattering optical element**. **a.** Micrometer-size optical field describing alphabet letter 'K' was constructed using the SOE. Regardless of the number and size of the wavefront segments, high-spatial-frequency components of the beams are readily accessed to form a micrometer-sized complex optical field. However, the background speckle field became significant when using a small number of optical modes. **b,** The ratio of the total intensity at the optimisation points to the

intensity of the background speckle field was measured as a function of number of segments. Red: optimised signal before recording. Blue: reconstructed signal using the SOE. Black dotted line: theoretically achievable value with perfect stationary circumstance.

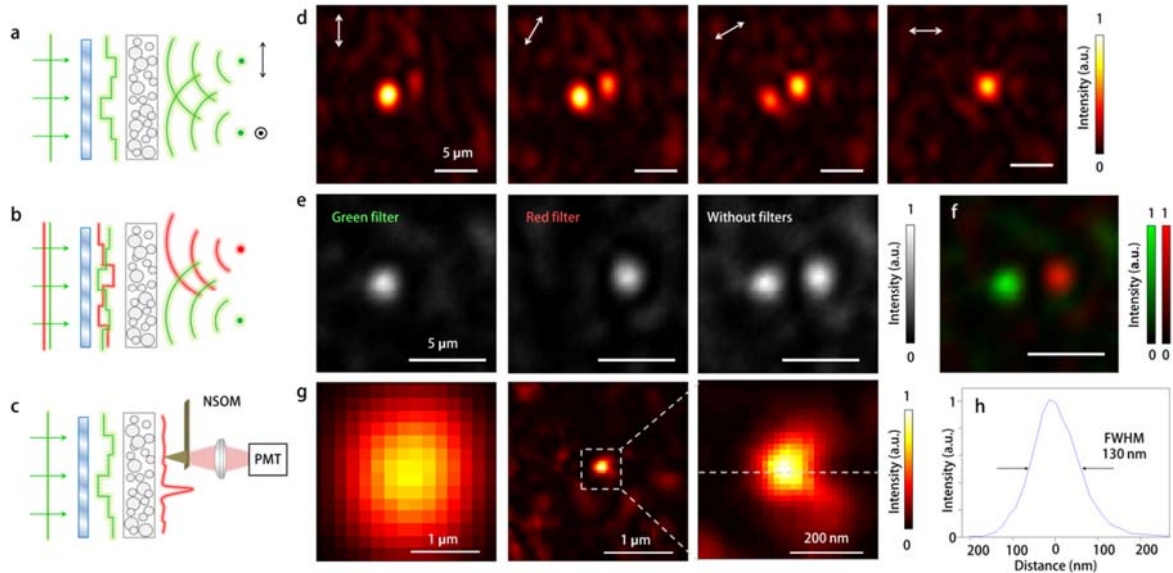

**Figure 5 | Access of polarisation states, spectral frequency components and near-field components of the optical field**. **a,** Access of polarisation states. Using turbid layers including randomly oriented birefringent liquid crystals, foci that have orthogonal polarisation states can be constructed. **d,** White arrows: orientation of analyzer. **b,** Spectral multiplexing. Foci having different wavelength component were constructed. **e**, The detection was performed by adding spectral filters in front of the CCD. **f**, Colored image constructed by using images from (**e**). **c,** Subwavelength light focusing. Using random nanoparticles as a turbid layer, a near-field focus was generated at the proximity of the turbid layer. The NSOM was used to optimise and raster-scan the near-field scattered signal from the turbid layer. **g**, (left) Far-field focus generated by the SOE. The image was captured by using CCD. (middle) Near-field focus generated by the SOE which was measured by using the NSOM. Image of the focus of was reacquired by dense raster-scanning (right).